\documentclass[aps,prd,preprint,groupedaddress,nofootinbib]{revtex4}

\usepackage{amssymb}
\begin{document}

\title{Minimal modifications of the primordial power spectrum from an
adiabatic short distance cutoff}   

\author{Jens C. Niemeyer}
 \email[]{jcn@mpa-garching.mpg.de}
\affiliation{Max-Planck-Institut f\"ur Astrophysik,
Karl-Schwarzschild-Str.~1, D-85748 Garching, Germany}
\author{Renaud Parentani}
\email[]{parenta@celfi.phys.univ-tours.fr}
\author{David Campo}
\email[]{campo@celfi.phys.univ-tours.fr}
\affiliation{Laboratoire de Math\'{e}matiques et Physique
Th\'{e}orique, CNRS UMR 6083,
Universit\'{e} de Tours, 37200 Tours, France}


\begin{abstract}
As a simple model for unknown Planck scale physics, we assume that the
quantum modes responsible for producing primordial curvature
perturbations during inflation are placed in their instantaneous
adiabatic vacuum when their proper momentum reaches a fixed high
energy scale $M$. The resulting power spectrum is derived and
presented in a form that exhibits the amplitude and frequency of the
superimposed oscillations in terms of $H/M$
and the slow roll parameter $\epsilon$. The amplitude of the
oscillations  is proportional to the third power of $H/M$.  We argue
that these small oscillations give the lower bound of the
modifications of the power spectrum if the notion of free mode
propagation ceases to exist above the critical energy scale $M$.
\end{abstract}


\maketitle

\section{Introduction}

Inflation, in addition to explaining the homogeneity and flatness of
the universe, provides a platform for computing the properties of the
small  inhomogeneities which give rise to the large scale  structures
\cite{L90,LL00}.  When initially in the vacuum, the quantized
fluctuations of the coupled system of metric and scalar inflaton field
produce a nearly scale invariant spectrum. Small deviations arise from
the time dependence of the  Hubble expansion rate during inflation
\cite{MFB92}.  The  predictions of the simplest inflationary scenarios
are in good agreement with recent analysis of high-resolution CMBR
maps.

However, these predictions are based on the incomplete framework of
semiclassical quantum field theory.  One particular aspect which has
been questioned lately is the choice of the ground state of the
quantum modes that generate the fluctuations. In the standard
derivation, the vacuum is imposed at  the beginning of inflation for
all quantum modes. During inflation,  the proper wavelengths of
these modes are redshifted by many orders of magnitude.  Given that
inflation lasts more than 70 e-folds, the initial state of the modes
responsible for the cosmological structures  observed today is
assigned at proper wavelengths shorter than the Planck length.
Moreover, in the semiclassical framework, one implicitly assumes that
their redshift to lower  momenta proceeds without interference from
Planck scale physics.  The validity of these assumptions is unclear;
this is known as the trans-Planckian question of inflationary
cosmology \cite{J00}.

The sensitivity of the predictions of inflation, in particular the
scale invariance of the power spectrum, with respect to
trans-Planckian physics has been probed in a number of different ways,
including nonlinear dispersion relations
\cite{MB00,BM00,BJM01,N00,NP01,S02}, string-inspired short  distance
cutoffs \cite{K00,KN01,Eea01,Eea01b,HS02}, and modifications derived
from non-commutative geometry \cite{CGS00,Lea02b,BH02}. Two aspects of
the discussion have become particularly  important: the question of
the leading order correction to the power spectrum in terms of the
ratio (called $\sigma$ in this work) of the Hubble parameter $H$ and
the scale of new physics $M$, and the existence of distinctive
features of short distance effects that would allow their unambiguous
detection in a general inflationary setting. As a candidate for the
latter, the potential violation of the scalar-tensor consistency
relation by Planck scale physics has been mentioned
\cite{HK01}. Unfortunately it requires the measurement of the tensor
perturbation power spectrum with great accuracy
\cite{Kea02}. Alternatively, if models for short distance physics
predict a unique pattern imprinted on the scalar power spectrum, this
alone might be sufficient for detection.

With regard to the amplitude of the correction, Ref.\cite{Kea02} has
pointed out that local  effective field theory generically allows only
$\sigma^2$ and smaller effects, whereas \cite{BM02} have argued that
$O(\sigma)$ effects are possible if adiabaticity of the mode evolution
is violated.  Danielsson  \cite{D02} has indeed found a linear
correction by imposing  that each quantum mode obeys a
positive frequency solution at a given energy scale $M$ (see Eq.(33)
in  \cite{Eea02}). This method has been re-analyzed in
Ref.\cite{Eea02} in a general inflationary background, confirming the
$O(\sigma)$-amplitude and demonstrating the oscillatory character of
the correction to the power spectrum. It has also been noted that some
classes of cutoff models predict corrections that depend on $M$ in more
complicated ways \cite{SW02}. 

In this work, we follow Danielsson's approach but we replace his
vacuum condition  by the instantaneous adiabatic vacuum at the scale
$M$.  The basic underlying physical assumption is that whatever
physics determines the dynamics at higher energies, it places the quantum
modes into their adiabatic vacuum at  some scale $M \lesssim m_{\rm
Planck}$, from where on they evolve freely and in the usual way,
i.e. without dispersion.  Owing to space-time expansion, this state is
not exactly identical to the asymptotic vacuum which is imposed in the
infinite past. This difference gives rise to a modification of the
power spectrum whose shape and amplitude we calculate both exactly and
perturbatively. In contrast with \cite{D02,Eea02}, we find a leading
correction that is cubic in $\sigma$, thus predicting a much smaller
modification of the spectrum.

The procedure of assigning the vacuum when the proper momentum of each
mode crosses $M$ can be considered a minimal model for short distance
effects since, in an expanding space-time, 
the adiabatic vacuum is the closest analogue to the
usual Minkowski vacuum.  Indeed, any other
choice of the initial state will  produce larger effects  unless
the amplitudes to find pairs of adiabatic quanta are fine tuned so as
to ``undo'' the Bogoliubov transformation with respect to the
asymptotic  vacuum, see the conclusions for more details.  Hence the
modification of the power spectrum we found  can be interpreted as the
minimal one if modes are ``created'' at the scale $M$.

Regarding the pattern that short distance physics
imprints on the microwave background, Easther et al. \cite{Eea01b}
have shown numerically for one specific model that nearly sinusoidal
oscillations in $\log(k)$ may be superimposed on the power
spectrum. This has been confirmed in \cite{Eea02} using the cutoff
criterion of \cite{D02}. We also find a distinctive oscillatory
pattern and present a simple expression for its oscillations in
wavenumber space. These oscillations arise through the wavenumber
dependence of $\sigma$, which is governed by the parameter $\epsilon$    
measuring the deviation of the inflationary background from de
Sitter space. Since in the slow-roll framework, $\epsilon$ can be
determined from the spectral index and the tensor/scalar ratio
\cite{K98b}, our model makes an explicit and  
simple prediction for the amplitude and shape of the correction. We
hope that this result will be useful in future studies of the
detectability of short distance effects in cosmological measurements.

\section{Adiabatic cutoff at the energy scale M}

The power spectrum of curvature perturbations produced during
inflation can be calculated from the dynamics of a minimally coupled,
massless scalar field in an FRW space time \cite{MFB92}. The $k$-mode
frequency component of this field obeys 
\begin{equation}
\label{omeg}
(\partial_\eta^2 + \omega^2_k(\eta) ) \phi_k = 0
\end{equation}
where $\eta$ is the conformal time. The time dependent frequency is 
given by
\begin{equation}
\label{omega}
\omega^2_k(\eta) =  k^2 - 
\frac{\partial_\eta^2 a(\eta)}{a(\eta)} \, .
\end{equation}

For simplicity, we restrict our discussion to power-law inflation with
\begin{equation}
a(\eta) = \left(\frac{H_0}{q} \, \eta \right)^q
\end{equation}
where $H_0$ is the value of the Hubble rate, $H = \partial_\eta a/a^2$,
evaluated at $\eta_0 = q/H_0 < 0$.
The parameter $q$ is related to the first slow roll parameter
$\epsilon$ by $q = -1/(1-\epsilon)$, where $\epsilon >0$ is defined as
\begin{equation}
\epsilon = - \frac{\partial_\eta H }{a H^2} = - \frac{\partial_t H }{ H^2}\, .
\end{equation}

Our basic assumption is that the unknown physics governed by quantum gravity
delivers each mode $k$ in the instantaneous
adiabatic vacuum at the time $\eta_k$ when the proper momentum $k/a(\eta)$ 
equals some high energy scale $M$. Below this scale, the 
propagation obeys Eq. (\ref{omeg}). This line of thought has been
originally developed in the context of Hawking radiation in \cite{J93}. 
The time $\eta_k$ is given by $k = a(\eta_k) M$ and can be written as
\begin{equation}
\eta_k = \frac{q}{k \sigma(\eta_k)}
\end{equation}
where we defined
\begin{equation}
\label{sigmadef}
\sigma(\eta) = \frac{H(\eta)}{M} \, .
\end{equation}
If we take $M$ to be of the order of the Planck mass, the amplitude of CMB
temperature fluctuations implies that $\sigma \lesssim 10^{-5}$. On
the other hand, certain strongly coupled M-theory scenarios allow somewhat
larger values for $\sigma$, reaching up to order unity
\cite{Eea01,Kea02}. We will nevertheless assume that $\sigma$ is a
small parameter for the purpose of determining the lowest order correction.
The time dependence of $\sigma$ can  be re-expressed in terms of $k$ using
$\eta_k = q/H_0\, \left(k/M\right)^{1/q}$ as
\begin{equation}
\label{sigmak}
\sigma_k \equiv \sigma( \eta_k)
= \frac{H_0}{M}\, \left(\frac{k}{M}\right)^{-\epsilon} =
\sigma_0 \, \left(\frac{k}{M}\right)^{-\epsilon}\,\,.
\end{equation}
This result is in agreement with Ref. \cite{Eea02} where a different
time coordinate was used. It should also be noted that $M$ itself can
slowly vary with $k$.
The formalism used in this work can readily handle this
extension as it is expressed entirely in terms of $\sigma_k$.  
  
To properly define the instantaneous adiabatic vacuum \cite{BD84} 
at time $\eta_k$ when $\partial_\eta \omega_k(\eta) \neq 0$ requires
some care. Our treatment closely follows Section 2.4 of \cite{MP98}
where more details can be found.  
We first require that the exact solution $\varphi_k^{\eta_k}(\eta)$
of Eq. (\ref{omeg}) is a positive frequency mode at $\eta = \eta_k$:
\begin{equation}
\label{varphidef}
i \partial_\eta \varphi_k^{\eta_k}(\eta)\vert_{\eta = \eta_k} =  
\omega_k(\eta_k) \,
\varphi_k^{\eta_k}(\eta_k)
\end{equation}
This solution is expressed as   
\begin{equation}
\label{phiwkb}
\varphi_k^{\eta_k}(\eta) = c_k^{\eta_k}(\eta) \varphi_k^{\rm WKB}(\eta) +
d_k^{\eta_k}(\eta) \left[\varphi_k^{\rm WKB}(\eta)\right]^\ast
\end{equation}
where $\varphi_k^{\rm WKB}(\eta)$ is the usual WKB solution of unit Wronskian:
\begin{equation}
\label{WKB}
\varphi_k^{\rm WKB}(\eta) =  \frac{1}{\sqrt{2 \omega_k(\eta)}}
 \exp\left(-i  \int^\eta_{\eta_k} \omega_k(\tilde \eta) d\tilde
 \eta\right) \, .
\end{equation}
Finally, we require that $i \partial_\eta \varphi_k^{\eta_k}
= \omega_k \left[ c_k^{\eta_k}
\varphi_k^{\rm WKB} - d_k^{\eta_k}(\varphi_k^{\rm WKB})^\ast\right]$
for all $\eta$. Then the Wronskian is conserved and Eq.~(\ref{omeg}) can be shown
to be equivalent to a set of coupled first order differential equations for
the coefficients $c_k^{\eta_k}(\eta)$ and $d_k^{\eta_k}(\eta)$
which govern non-adiabatic transitions (cf. \cite{MP98}, Eq.~47). 
The conditions that $\varphi_k^{\eta_k}(\eta)$ have 
unit Wronskian and be positive frequency at time $\eta_k$ translate into 
\begin{equation}
\label{ci}
c_k^{\eta_k}(\eta)\vert_{\eta = \eta_k} = 1 \,\,,\,\,\, 
d_k^{\eta_k}(\eta)\vert_{\eta = \eta_k} = 0\,\,.
\end{equation}
The instantaneous vacuum at $\eta_k$ is defined as the 
state annihilated by the ($\eta$-independent)
destruction operator associated with $\varphi_k^{\eta_k}(\eta)$. It
should be noted here that this construction corresponds to the zeroth
order in the adiabatic expansion \cite{BD84}. However, as can be
readily verified \cite{CNP02}, {\it all} higher order adiabatic
corrections give rise to only subleading corrections in the
$\sigma_0$-expansion of $\vert \beta_k^{\eta_k} \vert$ in
Eqs.~(\ref{betadS},\ref{betaappr}).  

The solution $\varphi_k^{\eta_k}(\eta)$
can be also expressed in terms of the asymptotic positive and
negative energy solutions. The positive frequency solution is 
\begin{equation}
\label{asymp}
\varphi_k^{-\infty} = 
\frac{1}{2} \sqrt{-\eta \pi}\,{\cal H}_{1/2 - q}(-k\eta)\,\,,
\end{equation}
where ${\cal H}_{1/2 - q}$ is the Hankel function of the first kind.
In the conventional derivation \cite{MFB92}, this solution is used to
define the vacuum state at the onset of inflation. The decomposition
of $\varphi_k^{\eta_k}$ reads 
\begin{equation}
\varphi_k^{\eta_k}(\eta) =  \alpha_k^{\eta_k} \varphi_k^{-\infty} +
\beta_k^{\eta_k} (\varphi_k^{-\infty})^\ast\,\,,
\end{equation}
where the Bogoliubov coefficients
$\alpha_k^{\eta_k}$ and $\beta_k^{\eta_k} $ are independent
of $\eta$. They are related to the coefficients of Eq. (\ref{phiwkb})
by $\alpha_k^{\eta_k}= c_k^{\eta_k}(-\infty)$ and
$\beta_k^{\eta_k}=d_k^{\eta_k}(-\infty)$ since $\varphi_k^{\rm WKB}$ is
asymptotically an exact positive frequency mode.
$\vert \beta_k^{\eta_k} \vert^2$
gives the mean occupation number of adiabatic instantaneous quanta
present in the asymptotic vacuum, or, equivalently, the number of
asymptotic quanta present in the instantaneous vacuum. 
We will use this latter interpretation to determine
the modifications of the power spectrum associated with
the replacement of the asymptotic vacuum by the instantaneous one.

\section{Modification of the power spectrum}

As explained in \cite{MFB92}, 
the power spectrum of scalar perturbations is governed by the 
norm of the solution evaluated at horizon crossing, i.e. 
when $k/a(\eta) = H(\eta)$. Thus, when working with
the instantaneous vacuum, it is given by
\begin{equation}
{\cal P}_{\cal R}^{\eta_k} = \frac{k^3}{2
\pi^2}\,\left\vert\frac{\varphi_k^{\eta_k}}{z}\right\vert_{k = a H}^2
\,\,,\,\,  z = - m_{\rm Planck} \, a \sqrt{\epsilon}\,\,,
\end{equation}
where
\begin{equation}
\vert\varphi_k^{\eta_k}(\eta)\vert^2 = \frac{\pi}{4}(-\eta) \left[(1 + 2
 \vert\beta_k^{\eta_k}\vert^2) \vert{\cal H}_{1/2 - q}
\vert^2 + 2 \Re\left\{
\alpha_k^{\eta_k}(\beta_k^{\eta_k})^\ast 
\left({\cal H}_{1/2 -q}
\right)^2 \right\}\right] \,.
\end{equation}
In the following, terms quadratic and higher order in
$\vert\beta_k^{\eta_k}\vert$ will be neglected assuming that
$\vert\beta_k^{\eta_k}\vert \ll 1$. 
The first order deviation from the usual 
spectrum ${\cal P}_{\cal R}^{-\infty}$
obtained with the asymptotic mode of Eq. (\ref{asymp}) is
\begin{equation}
\label{delta}
\delta_k \equiv  \frac{ {\cal P}^{\eta_k}_{\cal R}-{\cal P}^{-\infty}_{\cal R} }
{{\cal P}_{\cal R}^{-\infty}}
\simeq 2 \Re\left[ \frac{\beta_k^{\eta_k}}{\alpha_k^{\eta_k}} \,
\frac{({\cal H}_{1/2 - q})^\ast}
{{\cal H}_{1/2 - q}}\right]
\end{equation}
evaluated at $k/a = H$. Therefore, 
$\vert\delta_k\vert \sim 2 \vert\beta_k^{\eta_k}\vert$
provides the magnitude of the deviation from the standard result when
the adiabatic vacuum is imposed at $\eta_k$ instead of $\eta \to -\infty$.

What remains is to derive the coefficients $\alpha_k^{\eta_k}$ and
$\beta_k^{\eta_k}$. They are given by
\begin{eqnarray}
\left(\alpha_k^{\eta_k}\right)^\ast& 
= &(\varphi_k^{\eta_k})^\ast\, \tensor{i
\partial_\eta}\, \varphi_k^{-\infty}\, ,
\nonumber\\
\beta_k^{\eta_k}& =& \varphi_k^{\eta_k}\, \tensor{i
\partial_\eta}\, \varphi_k^{-\infty}\,\,.
\end{eqnarray}
Using Eqs.(\ref{varphidef}), (\ref{phiwkb}), (\ref{ci})
and evaluating (for simplicity) the r.h.s. at $\eta_k$, we get
\begin{equation}
\beta_k^{\eta_k} = i \left( \varphi_k^{\rm WKB} \,
\partial_\eta \varphi_k^{-\infty} + i \omega_k
\varphi_k^{\rm WKB} \, \varphi_k^{-\infty} \right)\vert_{\eta=\eta_k}\,\,,
\end{equation}
and a similar expression for $\alpha_k^{\eta_k}$. This yields
\begin{equation}
\label{coeffs}
\beta_k^{\eta_k} = \frac{-i}{4} 
\sqrt{\frac{\pi}{2 \vert \phi_k \vert}}
 \left[ x_k \left({\cal H}_{-1/2-q}( x_k) - {\cal
H}_{3/2-q}( x_k)\right) + \left(1 + i2 \phi_k\right){\cal
H}_{1/2-q}( x_k)\right] \, , 
\end{equation}
and $\alpha_k^{\eta_k} = \left[ \beta_k^{\eta_k}(-\phi_k)
\right]^\ast$, 
where 
$ x_k = -q/\sigma_k$, and, using Eq.(\ref{sigmak}),
\begin{eqnarray}
\label{phidef}
\phi_k \equiv \omega(\eta_k)\, \eta_k &=&
- \sqrt{x_k^2 - {q(q-1)}} 
\end{eqnarray}
Regarding the norm of the correction
$\vert\delta_k\vert \sim 2 \vert\beta_k^{\eta_k}\vert$,
we first give the exact expression for $\epsilon = 0$ (i.e., in de Sitter space)  
\begin{eqnarray}
\label{betadS}
\vert\beta_k^{\eta_k}\vert &=& \frac{1}{2}\left(\frac{2 - 2\sigma_0^2 -
 \sigma_0^4}{\sqrt{1 - 2 \sigma_0^2}} - 2\right)^{1/2}
\nonumber\\
&=& \frac{1}{2}  \sigma_0^3\left[ 1+ \frac{9}{8} \sigma_0^2 + 
O( \sigma_0^4) \right]\,\,.
\end{eqnarray}
In order to generalize this expression to $\epsilon \neq 0$, we use the asymptotic 
expansion of the Hankel function \cite{BO78}:
\begin{eqnarray}
{\cal H}_\nu(z)& = & \sqrt{\frac{2}{\pi z}}\,[ w_1(z) + i  w_2(z)]
\exp(iz - i\nu \pi/2 - i\pi/4) \nonumber\\
w_1(z) & \sim & \sum\limits_{n=0}^m (-1)^n c_{2n} z^{-2n}\nonumber\\
w_2(z) & \sim & \sum\limits_{n=0}^m (-1)^n c_{2n+1} z^{-2n-1}\nonumber\\
c_n & = & \frac{1}{8^n\,n!} \prod\limits_{j=1}^n \left(4 \nu^2 -
(2 j -1)^2\right)\,\,,\,\,c_0 = 1 \,\,,
\end{eqnarray}
where $m$ controls the order of the expansion. Expanding Eq.(\ref{coeffs})
around $\epsilon = \sigma = 0$ yields
\begin{equation}
\label{betaappr}
\vert\beta_k^{\eta_k}\vert = \left(\frac{1}{2} - \frac{3}{2}\epsilon
\left\{ \frac{1}{2}+ \ln(k/M) \right\} + O(\epsilon^2) \right)
\sigma_0^3 + O(\epsilon^3,\sigma_0^5)\,\,.
\end{equation}
Notice that this result is two powers of $\sigma$ smaller 
than the modification found by Danielsson \cite{D02}. The reason
is that the state he used is not the adiabatic vacuum. 
In fact, the Bogoliubov coefficients relating his modes 
to Eq.(\ref{phiwkb}) are $\beta_k= O(\sigma_k), \alpha_k \simeq 1$,
explaining the linear modification in $\sigma_k$.

Eq.(\ref{coeffs}) allows us to describe the oscillatory pattern
superimposed on the nearly scale invariant power spectrum. In
Eq.(\ref{delta}), the dependence on $k$ follows from
$\beta^{\eta_k}_k/\alpha^{\eta_k}_k$.  Then Eq. (\ref{coeffs}) shows
that $\delta$ is modulated by  $\cos(2x_k+ \chi_k)$ where $\chi_k$ 
is a slowly varying function of $k$. Rapid oscillations are generated
by the $k$-dependence of $\sigma_k= \sigma(\eta_k)$. Hence, any
deviation from de Sitter space will be reflected in an oscillatory
pattern  with frequency $(\epsilon/\sigma_0)\, \ln(k/M)$ if $\vert
\ln(k/M)\vert \ll \epsilon^{-1}$. A similar pattern has also been
found by  Easther et al. \cite{Eea01b,Eea02}. As $\epsilon$ can
independently be measured from the slope of the scalar power spectrum
and the tensor/scalar ratio, our result provides a simple prediction
for the signal of an adiabatic cutoff at the scale $M$.

\section{Discussion}

In spite of the smallness of $\delta$ in our model, this
result can be seen in an optimistic light. $\sigma$ may be larger than
the   normally assumed value of $10^{-5}$, allowing a potential
detection of short distance effects in CMBR measurements
\cite{Eea01b,Kea02}, especially in combination with a measurement of
the tensor spectrum. Moreover, it is by no means obvious that nature
chooses to place newly born modes in the instantaneous adiabatic
vacuum at $\eta_k$.  Indeed, when using a nonlinear dispersion
relation which modifies $\omega_k(\eta)$ for $k/a > M$, one might
obtain a state characterized by Bogoliubov coefficients $\alpha_k,
\beta_k$ which signal the departure from the adiabatic vacuum. In this
case, the leading correction will be given by $\vert \delta_k \vert
\simeq \vert  \beta_k/\alpha_k +
\beta^{\eta_k}_k/\alpha^{\eta_k}_k\vert$ in place of
Eq. (\ref{delta}). At this point, two remarks should be made.  First,
since $\vert \delta_k \vert $ is given by a sum of amplitudes and not
a sum of squares, one cannot exclude the possibility of a partial
cancellation of the two Bogoliubov transformations. However, an exact
cancellation is unlikely since $\beta_k$ and 
$\beta_k^{\eta_k}$ have different origins.  Secondly, one should keep
in mind that the coefficients $\beta_k$  must obey $\vert \beta_k
\vert \lesssim \sigma_k$  in order not to  violate the
hypothesis that the dominant contribution  to the Friedman equation be
the inflaton potential energy \cite{T00,NP01,S01}. Similar constraints
follow from the particle production rate in the late universe and
limits on the observed gamma-ray background \cite{ST02}. Whether there
exists a window in parameter space that allows the detection of a
high-energy cutoff in the CMBR remains an interesting question that,
in our opinion, cannot be fully answered at this point.

The simple prediction for the frequency of the superimposed
oscillation in the power spectrum given by Eq.(\ref{coeffs}) may allow
a systematic search for such features in cosmological data. If an
oscillation is detected whose frequency, for some $\sigma$, is
consistent with the value of $\epsilon$ inferred from the scalar
spectral index and the tensor/scalar ratio \cite{K98b}, it would
represent a strong indication of modified 
short distance physics. It is certainly worthwhile exploring the
parameter space that would make this detection feasible.

\begin{acknowledgments}
We thank Ulf Danielsson, Richard Easther, Brian Greene, Ted Jacobson,
Achim Kempf, Will Kinney, and Gary Shiu for helpful comments.
\end{acknowledgments}

\bibliography{early_universe}

\end{document}